\documentclass[preprint,nofootinbib,superscriptaddress,showpacs,amsmath]{revtex4}

\usepackage{graphicx}
\usepackage{dcolumn}
\usepackage{bm}

\usepackage{times}

\usepackage{graphicx}
\usepackage{dcolumn}
\usepackage{bm}

\begin{document}

\title{
Radiative corrections to the three-body region of the Dalitz plot of baryon semileptonic decays with angular correlation between polarized emitted baryons and charged leptons}

\author{
M.\ Neri
}
\affiliation{
Escuela Superior de F\'{\i}sica y Matem\'aticas del IPN, Apartado Postal 75-702, M\'exico, D.F.\ 07738, Mexico
}

\author{
J.\ J.\ Torres
}

\affiliation{
Escuela Superior de C\'omputo del IPN, Apartado Postal 75-702, M\'exico, D.F. 07738, Mexico
}

\author{
Rub\'en Flores-Mendieta
}

\affiliation{
Instituto de F{\'\i}sica, Universidad Aut\'onoma de San Luis Potos{\'\i}, \'Alvaro Obreg\'on 64, Zona Centro, San Luis Potos{\'\i}, S.L.P.\ 78000, Mexico
}

\author{
A.\ Mart{\'\i}nez
}

\affiliation{
Escuela Superior de F\'{\i}sica y Matem\'aticas del IPN, Apartado Postal 75-702, M\'exico, D.F.\ 07738, Mexico
}

\author{A.\ Garc{\'\i}a
}

\affiliation{
Departamento de F{\'\i}sica, Centro de Investigaci\'on y de Estudios Avanzados del IPN, Apartado Postal 14-740, M\'exico, D.F.\ 07000, Mexico
}

\date{\today}

\begin{abstract}
We have calculated the radiative corrections to the Dalitz plot of baryon semileptonic decays with angular correlation between polarized emitted baryons and charged leptons. This work covers both charged and neutral decaying baryons and is restricted to the so-called three-body region of the Dalitz plot. Also it is specialized at the center-of-mass frame of the emitted baryon. We have considered terms up to order $(\alpha/\pi)(q/M_1)^0$, where $q$ is the momentum transfer and $M_1$ is the mass of the decaying baryon, and neglected terms of order $(\alpha/\pi)(q/M_1)^n$ for $n \geq 1$. The expressions displayed are ready to obtain numerical results, suitable for model-independent experimental analyses.
\end{abstract}


\maketitle

\section{Introduction}

Currently, experiments on spin 1/2-baryon semileptonic decays (BSD), $A\to B\ell\nu_\ell$, where the polarization ${\hat {\mathbf s}_2}$ of the emitted baryon $B$ is observed are underway \cite{piccini}. The analysis of these experiments requires the inclusion of radiative corrections (RC) to the Dalitz plot when ${\hat {\mathbf s}_2}$ is nonzero. Our previous work \cite{neri07} does not cover this case. It is the purpose of this paper to produce such RC.

There are several requirements that must be met. In order to keep experimental analyses model independent it is necessary that RC are model independent themselves. There are many possible charge assignments to $A$ and $B$ and RC should be calculated so as to cover all the expected assignments. The charged lepton $\ell$ should be allowed to be an $e^\pm$, $\mu^\pm$, and even $\tau^\pm$ as the case may be. Since RC depend on the form factors present in the uncorrected decay amplitude, it is also necessary that they be cast into a form that can produce numerical results which are not compromised by fixing the form factors at prescribed values.

The model independence of RC is achieved by following  the generalization for hyperons \cite{rebeca} of the treatment of virtual RC in neutron beta decay \cite{sirlin} and of the application of the Low theorem \cite{low} to the bremsstrahlung RC developed in Ref.~\cite{chew}. There are six different charge assignments predicted by the light and heavy quark content of $A$ and $B$. To cover all these cases it is necessary to know only the RC to the neutral decaying baryon (NDB) $A^0\to B^+\ell^-\bar{\nu}_\ell$ and to the charged decaying baryon (CDB) $A^-\to B^0\ell^-\bar{\nu}_\ell$ cases. The other possibilities are obtained using the RC of the latter two \cite{mar02}. The cases $\ell=e^\pm$ , $\mu^\pm$, $\tau^\pm$ are included in the RC by keeping the mass $m$ of $\ell$ uncompromised all along the calculation. In order to produce numerical values of RC that are practical to use in the Monte Carlo simulation of an experimental analysis and that are not committed to fixed values of the form factors of the weak vertex, one can numerically calculate the RC to the coefficients of the quadratic products of form factors that appear in the theoretical differential decay rate of the decay being measured.

Since current experiments are medium-statistics (of the order of thousands of events) experiments and in order to keep the effort of calculating RC within convenient bounds, we shall consider contributions of order $(\alpha/\pi)(q/M_1)^n$, with $n=0$ only and neglect orders with $n=1$ and higher. Here $q$ is the four-momentum transfer and $M_1$ is the mass of $A$. Also we shall exhibit our results in a form where the integration over the real photon variables are ready to be performed numerically, except for the finite terms that accompany the infrared divergence of the bremsstrahlung RC which will be given analytically. The virtual RC will be given fully analytically. Our final result will be specialized to the center-of-mass frame of the emitted baryon $B$.

In Sec.\ II we introduce our notation and conventions and discuss in detail the boundaries of the Dalitz plot in the center-of-mass frame of $B$. We shall specialize our calculation to the three-body region of this plot. Section III is devoted to the model-independent calculation of virtual RC. We will see that they can be put formally in the same form of our previous work, although now they will be functions of the energies $E$ of $\ell$ and $E_1$ of $A$ in the center-of-mass frame of $B$. The rather long expressions containing the form factors that appear in these corrections are exhibited in  full in Appendix A. The bremsstrahlung RC are obtained in Sec.\ IV also in a model-independent form. However, the detailed discussion of its infrared divergence and the finite terms that accompany it is presented in Appendix B. In Sec.\ V we collect our results in a final form and we discuss their numerical use. We will cover the NDB and the CDB cases but we will exhibit only the calculation of the CDB case and limit ourselves to present the final results for the NDB case. Section VI is devoted to a brief discussion of our results. 

\section{Dalitz plot in the center-of-mass frame of the emitted baryon}

For definiteness, let us consider the BSD
\begin{equation}
A^- \rightarrow B^0 + \ell^- + \overline{\nu}_\ell. \label{eq:equno}
\end{equation}
The four-momenta and masses of the $A^-$,  $B^0$, $\ell^-$, and $\overline{\nu}_\ell,$ will be denoted by $p_1=(E_1,{\mathbf p}_1)$, $p_2=(E_2,{\mathbf p}_2)$, $l=(E,{\mathbf l})$, and $p_\nu=(E_\nu^0,{\mathbf p}_\nu)$, and by $M_1$, $M_2$, $m$, and $m_\nu$, respectively. The reference system we shall use is the center-of-mass frame of $B$. Accordingly, $E_2=M_2$ and ${\mathbf p}_2=\mathbf{0}$. It must be kept in mind that all other variables are referenced to this frame now. There should not arise any confusion with our previous work. A vanishing neutrino mass $m_\nu$ will be assumed. Additionally, the direction of a vector ${\mathbf p}$ will be denoted by a unit vector $\hat {\mathbf p}$ and whenever the expressions involved are not manifestly covariant, quantities like $p_1$, $l$, or $p_\nu$ will also denote the magnitudes of the corresponding three-momenta, unless stated otherwise.

The uncorrected transition amplitude $\mathsf{M}_0$ for process (\ref{eq:equno}) is given by the product of the matrix elements of the baryonic and leptonic currents, namely,
\begin{eqnarray}
\mathsf{M}_0 = \frac{G_V}{\sqrt 2} [\overline{u}_B(p_2) W_\mu(p_1,p_2) u_A(p_1)] [\overline{u}_\ell(l) O_\mu v_\nu(p_\nu)], \label{eq:eqdos}
\end{eqnarray}
where $u_A$, $u_B$, $u_\ell$, and $v_\nu$ are the Dirac spinors of the corresponding particles and $W_\mu$ is the weak interaction vertex given by
\begin{eqnarray}
W_\mu (p_1,p_2) & = & f_1(q^2) \gamma_\mu + f_2(q^2) \sigma_{\mu \nu} \frac{q_\nu}{M_1} + f_3(q^2) \frac{q_\mu}{M_1}
\nonumber \\
&  & \mbox{} + \left[g_1(q^2) \gamma_\mu + g_2(q^2) \sigma_{\mu \nu} \frac{q_\nu}{M_1} + g_3(q^2) \frac{q_\mu}{M_1} \right] \gamma_5. \label{eq:eqtres}
\end{eqnarray}
Here $O_\mu = \gamma_\mu (1+\gamma_5)$, $q\equiv p_1-p_2$ is the four-momentum transfer, and $f_i(q^2)$ and $g_i(q^2)$ are the conventional weak vector and axial-vector form factors, respectively, which are assumed to be real in this work. In Eq.~(\ref{eq:eqdos}) we have omitted the Cabibbo-Kobayashi-Maskawa factors. They should be inserted once decay (\ref{eq:equno}) is particularized.

To cover the observation of the polarization of $B$, its spinor is modified through the replacement
\begin{equation}
u_B(p_2) \rightarrow \Sigma(s_2) u_B(p_2), \label{eq:eqcinco}
\end{equation}
where $\Sigma(s_2)$, the spin projection operator, is given by
\begin{equation}
\Sigma(s_2) = \frac{1-\gamma_5 {\not \! s_2}}{2}, \label{eq:eqcuatro}
\end{equation}
and the polarization four-vector $s_2$ satisfies the relations $s_2 \cdot s_2 = -1$ and $s_2 \cdot p_2 = 0$. In the
center-of-mass frame of $B$, $s_2$ becomes the purely spatial unit vector $\hat {\mathbf s}_2$ which points along the spin direction. In the present calculation the results will be organized to explicitly exhibit the angular correlation ${\hat {\mathbf s}_2} \cdot \hat {\mathbf l}$.

\begin{figure}
\includegraphics{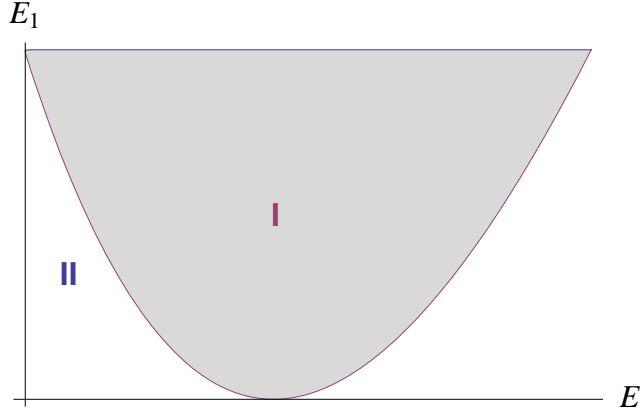}
\caption{\label{fig:kinem} Kinematical region as a function of $E$ and $E_1$ for baryon semileptonic decays. The areas $\mathsf{I}$ and $\mathsf{I+II}$ correspond to the Dalitz plots of the processes $A\rightarrow B+\ell + \overline{\nu}_\ell$ and $A \to B +\ell + \overline{\nu}_\ell + \gamma$, respectively.}
\end{figure}

Energy and momentum conservation determines the allowed kinematical region in the variables $E$ and $E_1$ for process (\ref{eq:equno}). This region, which is referred to as the Dalitz plot and is represented by the shadowed area depicted in Fig.~\ref{fig:kinem} and labeled as $\mathsf{I}$, is bounded in $E_1$ by
\begin{equation}
E_1^{\mathrm{min}} \leq E_1 \leq E_1^{\mathrm{max}}, \label{eq:eq46}
\end{equation}
where
\begin{equation}
E_1^{\mathrm{max,min}} = \frac{(M_2+E\mp l)^2 + M_1^2}{2(M_2+E \mp l)},
\end{equation}
while the charged lepton energy falls within the interval
\begin{equation}
m \leq E \leq E_m, \label{eq:eq47}
\end{equation}
where
\begin{equation}
E_m = \frac{M_1^2-M_2^2-m^2}{2M_2}. \label{eq:eq49}
\end{equation}

Similarly, area $\mathsf{II}$ in Fig.~\ref{fig:kinem} is bounded by
\begin{equation}
M_1 \leq E_1 \leq E_1^{\mathrm{min}}, \qquad \qquad m \leq E \leq E_c,
\end{equation}
where
\begin{equation}
E_c = \frac{(M_1-M_2)^2 + m^2}{2(M_1-M_2)}.
\end{equation}

The distinction between these two areas has important physical implications that should be clarified. Finding an event with energies $E$ and $E_1$ in area $\mathsf{II}$ demands the existence of a fourth particle which in our case will be a photon and will
carry away finite energy and momentum. In contrast, in area $\mathsf{I}$ this photon may or may not do so. In consequence, area $\mathsf{II}$ is exclusively a four-body region whereas area $\mathsf{I}$ is both a three- and a four-body region. We will refer loosely to areas $\mathsf{I}$ and $\mathsf{II}$ as the three- and four-body regions (TBR and FBR) of the Dalitz plot, respectively.

\section{\label{sec:vir}Virtual radiative corrections}

The method to calculate the virtual RC to the Dalitz plot of unpolarized and polarized decaying baryons has been discussed in detail in Refs.~\cite{rebeca} and \cite{sirlin}. It can be readily adapted to our case here of nonzero $\hat{\mathbf{s}}_2$, so only a few salient facts will be repeated now. The virtual RC can be separated into a model-independent part $\mathsf{M}_v$ which is finite and calculable and into a model-dependent one which contains the effects of the strong interactions and the intermediate vector boson. To order $(\alpha/\pi)(q/M_1)^0$, the latter amounts to two constants $(\alpha/\pi)c$ and $(\alpha/\pi)d$ which can be absorbed into $f_1$ and $g_1$ of $\mathsf{M}_0$, respectively, through the definition of effective form factors, hereafter referred to as $f_1^\prime$ and $g_1^\prime$. Thus, the decay amplitude $\mathsf{M}_V$  with virtual RC is given by
\begin{equation}
\mathsf{M}_V = \mathsf{M}_0^\prime + \mathsf{M}_v, \label{eq:eqseis}
\end{equation}
where
\begin{equation}
\mathsf{M}_v = \frac{\alpha}{2\pi} \left[ \mathsf{M}_0 \hat{\phi} + \mathsf{M}_{p_1} \hat{\phi}^\prime \right],
\label{eq:eqsiete}
\end{equation}
and
\begin{equation}
\mathsf{M}_{p_1} = \left(\frac{E}{m M_1} \right) \frac{G_V}{\sqrt 2} [\overline{u}_B W_\lambda u_A] [\overline{u}_\ell \! \! \not \! p_1 O_\lambda v_\nu]. \label{eq:eqdiez}
\end{equation}
The prime on $\mathsf{M}_0$ in Eq.~(\ref{eq:eqseis}) will be used as a reminder that the effective form factors
appear explicitly in this amplitude. Also, to order $(q/M_1)^0$ the amplitudes $\mathsf{M}_0$ and $\mathsf{M}_{p_1}$ in Eq.~(\ref{eq:eqsiete}) [but not in Eq.~(\ref{eq:eqseis})] are limited to contain only the leading form factors $f_1$ and $g_1$. The calculation of the model-independent functions $\hat{\phi}(E)$ and $\hat{\phi}^\prime(E)$ shows that they formally retain the same form given in previous work \cite{neri07}. The hats over them denote they are now given in the center-of-mass frame of the emitted baryon $B$. These functions read
\begin{eqnarray}
\hat{\phi}(E) & = & 2 \left[ \frac{1}{\beta} \tanh^{-1}\beta - 1 \right] \ln \left[ \frac{\lambda}{m} \right] - \frac{1}{\beta}(\tanh^{-1}\beta)^2 + \frac{1}{\beta} L \left[\frac{2\beta}{1+\beta}\right] + \frac{1}{\beta} \tanh^{-1}\beta - \frac{11}{8} \nonumber \\
&  & \mbox{} + \left\{ \begin{array}{ll}
\displaystyle \frac{\pi^2}{\beta} + \frac32 \ln \frac{M_2}{m} & \qquad (\textrm{NDB}) \\[0.6cm]
\displaystyle \frac32 \ln \frac{M_1}{m} & \qquad (\textrm{CDB})
\end{array} \right. \label{eq:eqocho}
\end{eqnarray}
and
\begin{equation}
\hat{\phi}^\prime (E) = \left[\beta - \frac{1}{\beta} \right] \tanh^{-1}\beta, \label{eq:eqnueve}
\end{equation}
where $\beta \equiv l/E$, $L$ is the Spence function, $\lambda$ is the infrared-divergent cutoff and with CDB and NDB we distinguish the results for the charged and neutral decaying baryon cases. The divergent term in Eq.~(\ref{eq:eqocho}) will be canceled by its counterpart in the bremsstrahlung contribution. 

At this point we can construct the Dalitz plot with virtual RC by leaving the energies $E$ and $E_1$ as the relevant variables in the differential decay rate for process (\ref{eq:equno}). After making the replacement (\ref{eq:eqcinco}) in (\ref{eq:eqseis}), squaring it, averaging over initial spins, summing over final spin states, and rearranging terms we can express the differential decay rate as \footnote{We shall also use hats over other expressions to emphasize that the center-of-mass frame of $B$ is being used.}
\begin{equation}
d\Gamma_V = d\hat{\Omega} \left\{ \hat{A}_0^\prime + \frac{\alpha}{\pi} (\hat{A}_1^\prime \hat{\phi} + \hat{A}_1^{\prime\prime} \hat{\phi}^\prime) - {\hat {\mathbf s}_2} \cdot \hat {\mathbf l} \left[ \hat{A}_0^{\prime\prime} + \frac{\alpha}{\pi} (\hat{A}_2^\prime \hat{\phi} + \hat{A}_2^{\prime\prime} \hat{\phi}^\prime) \right] \right\}, \label{eq:dGvir}
\end{equation}
where
\begin{equation}
d\hat{\Omega} = \frac12 \left[\frac{M_2}{M_1}\right]^3 \frac{G_V^2}{2} \frac{dEdE_1 d\Omega_\ell d\varphi_1}{(2\pi)^5} 2M_2.
\label{eq:dOm}
\end{equation}
Let us notice that this expression of $d\hat{\Omega}$ has some differences with respect to the one of previous work \cite{neri07}. These are the factor $1/2$, which results from averaging over the spin of the initial baryon, and the factor $(M_2/M_1)^3$, which arises out of the Lorentz transformation to the new reference frame. To recover the unpolarized decay rate one makes the factor 1/2 disappear by inserting in Eq.~(\ref{eq:eqdos}) the operator $\Sigma(-s_2) = (1+\gamma_5 {\not \! s_2})/2$ instead of (\ref{eq:eqcuatro}) and  adding the result to (\ref{eq:dGvir}).

The functions $\hat{A}_0^\prime$ and $\hat{A}_0^{\prime\prime}$, which emerge in the uncorrected amplitude $\mathsf{M}_0$, read
\begin{equation}
\hat{A}_0^\prime =  EE_\nu^0 \hat{Q}_1 - Ep_1 (p_1 - ly_0) \hat{Q}_2 - l(l-p_1y_0) \hat{Q}_3 - E_\nu^0p_1ly_0 \hat{Q}_4 + p_1^2ly_0(p_1-ly_0) \hat{Q}_5, \label{eq:a0p}
\end{equation}
and
\begin{equation}
\hat{A}_0^{\prime\prime} = Ep_1y_0 \hat{Q}_6 + El\hat{Q}_7, \label{eq:a0pp}
\end{equation}
where $y_0$ is defined as the scalar product ${\hat {\mathbf p}_1} \cdot \hat {\mathbf l}$ and can be expressed as
\begin{equation}
y_0 = \frac{p_1^2 + l^2 - {E_\nu^0}^2}{2p_1l}, \label{eq:y0}
\end{equation}
and also, by energy conservation, the neutrino energy $E_\nu^0$ is given by
\begin{equation}
E_\nu^0 = E_1 - E - M_2. \label{eq:enu0}
\end{equation}

The $\hat{Q}_i$ are new functions of the form factors and are listed in Appendix \ref{app:qs}. The hat is used to avoid confusing them with the ones of Ref.~\cite{neri07}. The functions $\hat{A}_1^\prime$, $\hat{A}_1^{\prime\prime}$, $\hat{A}_2^\prime$, and $\hat{A}_2^{\prime\prime}$ that emerge in these virtual RC read
\begin{equation}
\hat{A}_1^\prime = D_1EE_\nu^0 - D_2 l(l-p_1y_0), \label{eq:a1pr}
\end{equation}
\begin{equation}
\hat{A}_1^{\prime\prime} = \frac{M_2}{M_1} EE_\nu^0D_1, \label{eq:a1prpr}
\end{equation}
\begin{equation}
\hat{A}_2^\prime = E(l-p_1y_0) D_3 - lE_\nu^0D_4, \label{eq:a2pr}
\end{equation}
and
\begin{equation}
\hat{A}_2^{\prime\prime} = \frac{EM_2}{M_1}(l-p_1y_0) D_3, \label{eq:a2prpr}
\end{equation}
where the coefficients $D_i$ are quadratic functions of the effective form factors. Explicitly, they are
\begin{subequations}
\begin{eqnarray}
&  & D_1={f_1^\prime}^2 + 3{g_1^\prime}^2, \qquad \quad \quad D_2={f_1^\prime}^2 - {g_1^\prime}^2, \label{eq:d1d2} \\
&  & D_3= 2(-{g_1^\prime}^2+f_1^\prime g_1^\prime), \qquad D_4=2({g_1^\prime}^2 + f_1^\prime g_1^\prime). \label{eq:d3d4}
\end{eqnarray}
\end{subequations}
Here we use also the effective form factors, so that our result is uniformly expressed. This is a rearrangement of second order in $\alpha /\pi$, which we are free to make within our approximations.  

To stress the parallelism with our previous work we have used the same notation, but there should arise no confusion. The expressions given here apply to the present case only.

\section{\label{sec:bre}Bremsstrahlung radiative corrections}

To obtain the bremsstrahlung RC we have to consider the four-body decay
\begin{equation}
A^- \to B^0 +\ell^- + \overline{\nu}_\ell + \gamma, \label{eq:eq36}
\end{equation}
where $\gamma$ represents a massive  photon with four-momentum $k=(\omega,\mathbf{k})$ and $k^2=\lambda^2$. To obtain the bremsstrahlung RC in a model-independent way we shall use the Low theorem \cite{low,chew}, which asserts that the radiative amplitudes of order $1/k$ and $(k)^0$ can be determined in terms of the nonradiative amplitude without further structure dependence. Then, we can express the bremsstrahlung amplitude $\mathsf{M}_B$ as
\begin{equation}
\mathsf{M}_B=\mathsf{M}_{B1}+\mathsf{M}_{B2}+\mathsf{M}_{B3}, \label{eq:MB}
\end{equation}
with
\begin{equation}
\mathsf{M}_{B1} = e\mathsf{M}_0 \left[\frac{2l\cdot\epsilon}{2l\cdot k+\lambda^2} + \frac{2p_1\cdot\epsilon}{\lambda^2-2p_1\cdot k} \right], \label{eq:eq39}
\end{equation}
and
\begin{equation}
\mathsf{M}_{B2} = \frac{eG_V}{\sqrt 2} \epsilon_\mu [\overline{u}_B W_\lambda u_A] [\overline{u}_\ell \frac{\gamma_\mu \!\! \not \! k}{2l\cdot k+\lambda^2} O_\lambda v_\nu]. \label{eq:eq40}
\end{equation}
$\mathsf{M}_{B1}$ contains terms of order $1/k$ and $\mathsf{M}_{B2}$ contains terms of order $(k)^0$. Although $\mathsf{M}_{B3}$ contains also some terms of order $(k)^0$, it can be ignored because its contribution to the decay rate is of order $q/M_1$ \cite{neri07} so we do not need its explicit form. The infrared-divergent terms are all contained in $\mathsf{M}_{B1}$.

Next, we have to replace Eq.~(\ref{eq:eqcinco}) in Eq.~(\ref{eq:MB}), square the resulting $\mathsf{M}_B$, average over the initial spins and sum over the final spins and over the photon polarization. To perform the latter sum, we proceed in two ways. First, we can use the rule of Coester \cite{coester} to account for the longitudinal degree of polarization of the photon, namely,
\begin{equation}
\sum_\epsilon (\epsilon \cdot a)(\epsilon \cdot b) = \mathbf{a} \cdot \mathbf{b} - \frac{(\mathbf{a} \cdot \mathbf{k}) (\mathbf{b} \cdot \mathbf{k})}{\omega^2} \label{eq:coester}
\end{equation}
where $\omega^2=k^2+\lambda^2$ (here $k$ is the magnitude of $\mathbf{k}$) and $a=(a_0,\mathbf{a})$ and $b=(b_0,\mathbf{b})$ are arbitrary 4-vectors. Second, the infrared-convergent contributions can be calculated using the usual summation over the photon polarization, namely, $\sum_\epsilon(\epsilon \cdot a) (\epsilon \cdot b) = -a\cdot b$, with $\omega =k$.

After a standard calculation, the bremsstrahlung differential decay rate $d\Gamma_B$ corresponding to the TBR becomes
\begin{equation}
d\Gamma_B = d\Gamma_B^\prime - d\Gamma_B^{(s)}, \label{eq:dGb}
\end{equation}
where $d\Gamma_B^\prime$ denotes half of the unpolarized decay rate whereas $d\Gamma_B^{(s)}$ contains the spin of the emitted baryon. Explicitly, using the $D_i$ of Eqs.~(\ref{eq:d1d2}) and (\ref{eq:d3d4}) \footnote{For the sake of the uniformity of our results, the effective form factors $f_1^\prime$ and $g_1^\prime$ may be used here, too. Again, this amounts to a rearrangement of order $(\alpha/\pi)^2$, valid within our approximations.}, they are,
\begin{eqnarray}
d\Gamma_B^\prime & = & \frac12 \frac{\alpha}{\pi} \frac{G_V^2}{2} \frac{2M_2}{(2\pi)^6} \left[\frac{M_2}{M_1}\right]^3 \frac{d^3p_1}{E_1} \frac{d^3l}{E} \frac{d^3k}{\omega} \frac{d^3p_\nu}{E_\nu} \delta^4 (p_1-p_2-l-p_\nu-k) \nonumber \\
&  & \mbox{} \times \left\{ \frac{\beta^2 (1-k^2x^2/\omega^2)}{(\omega-\beta kx)^2} (D_1EE_\nu+D_2 \mathbf{l}
\cdot \mathbf{p}_\nu) \right. \nonumber \\
&  & \mbox{} + \frac{1}{E(\omega-\beta kx)} \left[D_1E_\nu \left(\omega+2E-\frac{m^2\omega}{E(\omega -\beta kx)} - \frac{E(\omega - \beta kx)}{\omega}\right) \right. \nonumber \\
&  & \mbox{} \mbox{} \left. \left. + D_2\left[ \mathbf{k} \cdot \mathbf{p}_\nu \left( 1-\frac{m^2}{E(\omega-\beta kx)} + \frac{E}{\omega} \right) + \mathbf{l} \cdot \mathbf{p}_\nu\right] \right] \right\}, \label{eq:Dgbp}
\end{eqnarray}
and
\begin{eqnarray}
d\Gamma_B^{(s)} & = & -\frac12 \frac{\alpha}{\pi} \frac{G_V^2}{2} \frac{2M_2}{(2\pi)^6} \left[\frac{M_2}{M_1}\right]^3 \frac{d^3p_1}{E_1} \frac{d^3l}{E} \frac{d^3k}{\omega} \frac{d^3p_\nu}{E_\nu} \delta^4(p_1-p_2-l-p_\nu-k) \,
({\hat {\mathbf s}_2} \cdot {\hat {\mathbf l}}) \nonumber \\
&  & \mbox{} \times \left\{ \frac{\beta^2(1-k^2x^2/\omega^2)}{(\omega-\beta kx)^2} (D_3E {\hat {\mathbf l}} \cdot
\mathbf{p}_\nu + D_4 E_\nu l) \right. \nonumber \\
&  & \mbox{} + \frac{1}{E(\omega-\beta kx)} \left[D_3 {\hat {\mathbf l}} \cdot {\mathbf p}_\nu \left(\omega+2E-\frac{m^2\omega}{E(\omega-\beta kx)} - \frac{E(\omega-\beta kx)}{\omega} \right) \right. \nonumber \\
&  & \mbox{} \left. \left. + D_4E_\nu\left[ \mathbf{k}\cdot {\hat {\mathbf l}} \left(1-\frac{m^2}{E(\omega-\beta kx)} + \frac{E}{\omega} \right) + l \right] \right] \right\}.  \label{eq:Dgbs}
\end{eqnarray}
The above expression of $d\Gamma_B^{(s)}$ is specialized to the angular correlation ${\hat {\mathbf s}}_2 \cdot {\hat {\mathbf l}}$. To achieve this we have used the replacement ${\hat {\mathbf s}}_2 \cdot {\mathbf p} \to ( {\hat {\mathbf s}}_2 \cdot {\hat {\mathbf l}}) ( {\hat {\mathbf l}} \cdot \mathbf{p})$, with ${\mathbf p} = {\mathbf p}_1, {\mathbf k}, {\mathbf p}_\nu$, which is valid over the Dalitz plot after all other variables are integrated. Both expressions (\ref{eq:Dgbp}) and (\ref{eq:Dgbs}) contain the infrared divergence in their first summand within the curly brackets, which we analyze in detail in Appendix \ref{app:ir}. The $\lambda^2$ that appears in Eq.~(\ref{eq:eq39}) contributes in Eqs.~(\ref{eq:Dgbp}) and (\ref{eq:Dgbs}) with linear and higher powers, which will become zero in the $\lambda^2 \to 0$ limit.

The integration over the neutrino variables with $\delta^3({\mathbf p}_1-{\mathbf l}-{\mathbf p}_\nu-{\mathbf k})$ in Eqs.~(\ref{eq:Dgbp}) and (\ref{eq:Dgbs}) is trivial, but leaves a nontrivial argument inside the last $\delta(E_1-M_2-E-E_\nu-\omega)$, which allows one to integrate over the photon momentum. Without further ado, the resulting expressions can be cast into
\begin{equation}
d\Gamma_B^\prime = \frac{\alpha}{\pi} d\hat{\Omega} \left[ \hat{A}_1^\prime \hat{I}_0 + (\rho_1+\rho_1^\prime) D_1 + (\rho_2+\rho_2^\prime) D_2 \right],  \label{eq:dGbp}
\end{equation}
and
\begin{equation}
d\Gamma_B^{(s)} = \frac{\alpha}{\pi} d\hat{\Omega} \, {\hat {\mathbf s}}_2 \cdot {\hat {\mathbf l}} \left[ \hat{A}_2^\prime \hat{I}_0 + (\rho_3+\rho_3^\prime) D_3 + (\rho_4+\rho_4^\prime) D_4 \right], \label{eq:dGbs}
\end{equation}
where $d\hat{\Omega}$, $\hat{A}_1^\prime$, $\hat{A}_2^\prime$, $D_1$, $D_2$, $D_3$, and $D_4$ were already defined.  The integrals over the angular variables of the photon, $x={\hat {\mathbf k}} \cdot {\hat {\mathbf l}}$ and $\varphi_k$, and over $y={\hat {\mathbf p}_1} \cdot {\hat {\mathbf l}}$ are left to be performed numerically, except in $\hat{I}_0$ which contains the infrared divergence and is obtained analytically. Its result is found in Appendix \ref{app:ir}. The functions $\rho_1$, $\rho_1^\prime$, $\rho_2$, $\rho_2^\prime$, $\rho_3$, $\rho_3^\prime$, $\rho_4$, and $\rho_4^\prime$ thus read
\begin{eqnarray}
\rho_1 & = & \frac{\beta^2p_1lE}{4\pi} \int_1^{y_0} dy \int_{-1}^1 dx \int_0^{2\pi} \frac{d\varphi_k}{D} \frac{1-x^2}{(1-\beta x)^2}, \label{eq:rho1} \\
\rho_1^\prime & = & -\frac{p_1l}{4\pi} \int_1^{y_0} dy \int_{-1}^1 dx \int_0^{2\pi} \frac{d\varphi_k}{D} \frac{E_\nu}{1-\beta x} \left[\frac{\beta^2(1-x^2)}{1-\beta x} + \frac{\omega}{E}\right], \\
\rho_2 & = & -\frac{\beta^2p_1l}{4\pi} \int_1^{y_0} dy \int_{-1}^1 dx \int_0^{2\pi} d\varphi_k \frac{1-x^2}{(1-\beta x)^2} \left[ 1-\frac{lx}{D} \right],\label{eq:rho2} \\
\rho_2^\prime & = & -\frac{p_1l}{4\pi} \int_1^{y_0} dy \int_{-1}^1 dx \int_0^{2\pi} \frac{d\varphi_k}{D} \frac{1}{1-\beta x} \left[ {\hat {\mathbf k}} \cdot {\mathbf p}_\nu \left(1-\frac{1-\beta^2}{1-\beta x} + \frac{\omega}{E} \right) + \beta {\hat {\mathbf l}} \cdot {\mathbf p}_\nu \right], \\
\rho_3 & = & \frac{\beta p_1l}{4\pi} \int_1^{y_0} dy \int_{-1}^1 dx \int_0^{2\pi} d\varphi_k \frac{1-x^2}{(1-\beta x)^2} \left[1-\frac{lx}{D}\right] \\
& = & \mbox{} -\frac{\rho_2}{\beta}, \label{eq:rho3} \\
\rho_3^\prime & = & -\frac{p_1l}{4\pi} \int_1^{y_0} dy \int_{-1}^1 dx \int_0^{2\pi} \frac{d\varphi_k}{D}
\frac{l-p_1y+\omega x}{1-\beta x} \left[\frac{\beta^2(1-x^2)}{1-\beta x} + \frac{\omega}{E} \right], \\
\rho_4 & = & -\frac{\beta^2p_1l^2}{4\pi} \int_1^{y_0} dy \int_{-1}^1 dx \int_0^{2\pi} \frac{d\varphi_k}{D} \frac{1-x^2}{(1-\beta x)^2} \\
& = & \mbox{} -\beta\rho_1, \\
\rho_4^\prime & = & \frac{p_1l}{4\pi} \int_1^{y_0} dy \int_{-1}^1 dx \int_0^{2\pi} d\varphi_k \frac{E_\nu}{D} \frac{1}{1-\beta x} \left[ \beta + x\left(1-\frac{1-\beta^2}{1-\beta x} + \frac{\omega}{E} \right) \right], \label{eq:rho4p}
\end{eqnarray}
where
\begin{equation}
\omega = \frac{p_1l(y-y_0)}{D},
\end{equation}
with
\begin{equation}
D = E_\nu^0 - ( {\mathbf p}_1 - {\mathbf l}) \cdot {\hat {\mathbf k}}.  \label{eq:de}
\end{equation}

We have obtained the bremsstrahlung RC to the differential decay rate to order $(\alpha/\pi)(q/M_1)^0$. In the next section we present the total differential decay rate by gathering both virtual and bremsstrahlung contributions together.

\section{\label{sec:numerical} Final results and numerical form of the radiative corrections}

We have reached our first goal: To obtain the complete RC to the Dalitz plot with the ${\hat {\mathbf s}}_2 \cdot {\hat {\mathbf l}}$ correlation to order $(\alpha/\pi)(q/M_1)^0$ restricted to the TBR. The final result is obtained by summing the virtual RC, Eq.~(\ref{eq:dGvir}), and the bremsstrahlung RC, Eq.~(\ref{eq:dGb}). The latter is obtained when Eqs.~(\ref{eq:dGbp}) and (\ref{eq:dGbs}) are put together. Thus
\begin{equation}
d\Gamma = d\Gamma_V + d\Gamma_B.  \label{eq:dg}
\end{equation}

We can rearrange this expression into a simple form as
\begin{equation}
d\Gamma = d\hat{\Omega} \left\{ \hat{A}_0^\prime + \frac{\alpha}{\pi} \hat{ \Theta}_I - {\hat {\mathbf s}}_2 \cdot {\hat {\mathbf l}} \left[ \hat{A}_0^{\prime\prime} + \frac{\alpha}{\pi} \hat{\Theta}_{II} \right] \right\}, \label{eq:dGt}
\end{equation}
with
\begin{equation}
\hat{\Theta}_I = \hat{A}_1^\prime (\hat{\phi}+\hat{I}_0) + \hat{A}_1^{\prime\prime} \hat{\phi}^\prime + (\rho_1+\rho_1^\prime) D_1 + (\rho_2+\rho_2^\prime) D_2, \label{eq:teta1}
\end{equation}
and
\begin{equation}
\hat{\Theta}_{II} = \hat{A}_2^\prime (\hat{\phi}+\hat{I}_0) + \hat{A}_2^{\prime\prime}\hat{\phi}^\prime + (\rho_3+\rho_3^\prime) D_3 + (\rho_4+\rho_4^\prime) D_4, \label{eq:teta2}
\end{equation}
where all the ingredients are given in the previous sections.

Notice that the only difference between the NDB and the CDB cases is found in the function  $\hat{\phi}(E)$, Eq.~(\ref{eq:eqocho}). The bremsstrahlung RC does not make any difference between both cases because of the order of approximation in this work.

We now come to our second goal in this paper. This Eq.~(\ref{eq:dGt}) has triple integrals over some angular variables ready to be performed numerically. It requires that the numerical integrals in the RC be calculated within a Monte Carlo simulation every time $E$, $E_1$, and the form factors are varied, a task that represents a non-negligible computer effort. We shall now discuss a second form of the RC that should be more practical to use.

For fixed values of $E$ and $E_1$, Eqs.~(\ref{eq:teta1}) and (\ref{eq:teta2}) take the form
\begin{equation}
\hat{\Theta}_{r} = a_{r}f_1^2+b_{r}f_1g_1+c_{r}g_1^2, \label{eq:tetam}
\end{equation}
because they are quadratic in the form factors. The subindex $r$ takes the values $r=I$, $II$. The second form of RC we propose consists of calculating arrays of the $a_r$, $b_r$, and $c_r$ coefficients determined at fixed values of $(E,E_1)$ and that these pairs of $(E,E_1)$ cover a lattice of points on the Dalitz plot.

To calculate the coefficients $a_r$, $b_r$ and $c_r$ it is not necessary to rearrange our final results to take the form (\ref{eq:tetam}). One can calculate them following a systematic procedure. One chooses fixed $(E,E_1)$ points. Then one fixes $f_1=1$ and $g_1=0$, and obtains $a_{r}$, one repeats this calculation for $g_1=1$, $f_1=0$, to obtain $c_{r}$. Next, one repeats the calculation with $f_1=1$, $g_1=1$, and from these results one subtracts $a_{r}$ and $c_{r}$, this way one obtains the coefficient $b_{r}$. The arrays of these coefficients should be fed into the Monte Carlo simulation. Within this simulation the repetitive triple integrations are reduced into a form of matrix multiplication.

We may close this section by stressing that none of the forms of our RC results is compromised to fixing from the outset values for the form factors when such RC are applied in a Monte Carlo simulation. 

\section{Discussions}

Our final result for the RC to the Dalitz plot of BSD with the angular correlation ${\hat {\mathbf s}}_2 \cdot {\hat {\mathbf l}}$ between the polarization of the emitted baryon and the direction of the charged lepton is given in Eq.~(\ref{eq:dGt}). It is valid to order $(\alpha/\pi)(q/M_1)^0$ and it covers the TBR of this plot. It meets the requirements discussed in the introductory section, namely, it is model independent, it can be used in all charged assignments in different BSD, the charged lepton may be $e^\pm$, $\mu^\pm$, or $\tau^\pm$, and it is not compromised to fixed values of the form factors of the uncorrected decay amplitude. The finite terms that accompany the infrared divergence in the bremsstrahlung RC are given in analytical form. The other terms in this correction are presented  with triple integrations ready to be performed.

This result may be used in a Monte Carlo simulation of an experimental analysis. However, performing the triple integrations every time $E$, $E_1$, and the form factors are varied may represent a very heavy computer effort. A more practical use of our result is through Eq.~(\ref{eq:tetam}). Numerical arrays of the RC to the coefficients of the quadratic products of form factors may be first obtained at a lattice of points $(E,E_1)$ covering the Dalitz plot and afterwards be fed in the Monte Carlo simulation. The computer effort within it would then be reduced to a sort of matrix multiplication. 

The procedure of this paper may be followed in the future to extend the calculation of RC to BSD with the observation of polarization of the emitted baryon to cover the angular correlation ${\hat {\mathbf s}}_2 \cdot {\hat {\mathbf p}_1}$ and the FBR. The precision of RC may be improved, while still preserving their model independence, by including terms of order $(\alpha/\pi)(q/M_1)^n$ with $n=1$. It may be the case that experimental analyses should be limited to the center-of-mass frame of the decaying baryon $A$. Our results should be then adapted to this frame. Each one of these possibilities requires further serious efforts. They should be attempted as the need for them arises.

\acknowledgments

The authors are grateful to Consejo Nacional de Ciencia y Tecnolog{\'\i}a (Mexico) for partial support. J.~J.~T.\ and A.~M.\ were partially supported by Comisi\'on de Operaci\'on y Fomento de Actividades Acad\'emicas (Instituto Polit\'ecnico Nacional). R.~F.-M.\ was also partially supported by Fondo de Apoyo a la Investigaci\'on (Universidad Aut\'onoma de San Luis Potos{\'\i}).

\appendix

\section{\label{app:qs}The $\hat{Q}_i$ coefficients}

The $\hat{Q}_i$ factors contained in Eqs.~(\ref{eq:a0p}) and (\ref{eq:a0pp}) are quadratic functions of the form factors. For the spin-independent contribution they read
\begin{equation}
\hat{Q}_1 = \frac{E_1}{M_2} (F_1^2+G_1^2-2F_1G_1+F_1F_2-G_1G_2) + (1-\beta^2)\frac{E}{M_2}(F_1F_3-G_1G_3) + E_1 \hat{Q}_2 - \hat{Q}_3,
\end{equation}
\begin{equation}
M_1 \hat{Q}_2 = \frac{4M_1}{M_2}F_1G_1 + (1-\beta^2)\frac{E}{M_2}\left[ F_1F_3+G_1G_3 + \frac{E_1+M_1}{M_1}F_2F_3 + \frac{E_1-M_1}{M_1}G_2G_3 \right] + M_1 \hat{Q}_4,
\end{equation}
\begin{eqnarray}
\hat{Q}_3 & = & \frac{M_1}{M_2}(F_1^2-G_1^2) - \frac12 (1-\beta^2) \frac{E^2}{M_1^2} \left[ \frac{E_1+M_1}{M_2}F_3^2 + \frac{E_1-M_1}{M_2} G_3^2 \right] \nonumber \\
&  & \mbox{} + \frac{E_1}{M_2} (F_1F_2 - G_1G_2)  + \frac{M_1^2}{2}\hat{Q}_5,
\end{eqnarray}
\begin{equation}
\hat{Q}_4 = \frac{1}{M_2} (F_1^2 + G_1^2 - 2F_1G_1 + F_1F_2 - G_1G_2) + E_1\hat{Q}_5,
\end{equation}
\begin{equation}
M_1^2\hat{Q}_5 = \frac{2M_1}{M_2}(F_1F_2 + G_1G_2) + \frac{E_1+M_1}{M_2}F_2^2 + \frac{E_1-M_1}{M_2} G_2^2,
\end{equation}
whereas for the spin-dependent contribution they read
\begin{eqnarray}
\hat{Q}_6 & = & \left[\frac{E_1-M_1-\beta p_1 y_0}{M_2}\right] F_1^2 + \left[\frac{E_1+M_1-\beta p_1 y_0}{M_2}\right] G_1^2 - \frac{\beta p_1 y_0}{M_1} (F_1F_2 + G_1 G_2) \nonumber \\
&  & \mbox{} - \left[\frac{E_1-\beta p_1 y_0}{M_2}\right]2 F_1G_1 + \left[ 1+(1-\beta^2)\frac{E}{M_2}\left(1-\frac{E}{M_1}\right) - \frac{M_2+2E-\beta p_1y_0}{M_1}\right] F_1G_2 \nonumber \\
&  & \mbox{} - \left[1+(1-\beta^2)\frac{E}{M_2}\left(1+\frac{E}{M_1}\right) + \frac{M_2+2E-\beta p_1y_0}{M_1}\right] F_2G_1 \nonumber \\
&  & \mbox{} + (1-\beta^2)\frac{E}{M_2}\left[\left(1 - \frac{M_2}{M_1}-\frac{E}{M_1}\right) F_1G_3 - \left(1 + \frac{M_2}{M_1}+\frac{E}{M_1} \right) F_3G_1 \right]\nonumber \\
&  & \mbox{} + \left[ 1 + (1-\beta^2)\frac{E}{M_2} + \frac{E_1-\beta p_1y_0}{M_2} \left( 1 - \frac{2(E_1E-p_1ly_0)+2M_2E_1}{M_1^2}\right) \right] F_2G_2 \nonumber \\
&  & \mbox{} + (1-\beta^2)\frac{E}{M_1}\left[\frac{M_1}{M_2}-\frac{E_1}{M_1}-\frac{E}{M_1}\frac{E_1-\beta p_1y_0}{M_2} \right] (F_2G_3 + F_3G_2) \nonumber \\
&  & \mbox{} + (1-\beta^2) \frac{E^2}{M_1^2}\left[ \frac{E_\nu^0-\beta(p_1 y_0-l)}{M_2} \right]F_3G_3, \label{eq:q6}
\end{eqnarray}
and
\begin{eqnarray}
\hat{Q}_7 & = & \left[1+\frac{M_1}{M_2} \right] \left[\frac{E_1-M_1}{E} \right] F_1^2 + \left[1-\frac{M_1}{M_2} \right] \left[\frac{E_1+M_1}{E} \right] G_1^2 + \frac{p_1^2}{M_1 E} (F_1F_2+G_1G_2) \nonumber \\
&  & \mbox{} + \left[(1-\beta^2)\frac{E}{M_2}-\frac{E_\nu^0-E}{E}\right] \left[ 2F_1G_1+\frac{E_1-M_1}{M_1} F_1G_2 + \frac{E_1+M_1}{M_1} F_2G_1 \right] \nonumber \\
&  & \mbox{} + (1-\beta^2)\frac{E}{M_2} \left[\frac{E_1-M_1}{M_1} F_1G_3 + \frac{E_1+M_1}{M_1} F_3G_1 \right].
\end{eqnarray}

In these expressions we have used the form factors $F_i$ and $G_i$, which read
\begin{eqnarray}
&  & F_1 = f_1^\prime + \left(1+\frac{M_2}{M_1} \right) f_2, \qquad F_2 = -2 f_2, \qquad F_3=f_2+f_3,\nonumber \\
&  & G_1 = g_1^\prime - \left(1-\frac{M_2}{M_1} \right) g_2, \qquad G_2 = -2 g_2, \qquad G_3=g_2+g_3.\nonumber
\end{eqnarray}

\section{\label{app:ir}Extraction of the infrared divergence}

Here we discuss the procedure we followed to identify and isolate the infrared divergence contained in the $\hat{I}_{0}$ function introduced in Eqs.~(\ref{eq:dGbp}) and (\ref{eq:dGbs}). We only show how the infrared-divergent term is calculated in the spin-independent contribution, because in the spin-dependent one the procedure is analogous. The term where the infrared divergence is contained is
\begin{equation}
\sum_{\mathrm{spins},\epsilon} \left\vert \mathsf{M}_{B1}^\prime \right\vert^2 = \frac{e^2G_V^2}{2} \sum_\epsilon \left[\frac{2l\cdot\epsilon}{2l\cdot k+\lambda^2} + \frac{2p_1\cdot\epsilon}{\lambda^2-2p_1\cdot k} \right]^2 \frac{2M_2}{M_1mm_\nu}(N+N^\prime), \label{eq:app1}
\end{equation}
where the factors $N$ and $N^\prime$ are
\begin{equation}
N=D_1EE_\nu^0-D_2l(l-p_1y), \quad \quad N^\prime = -\omega (ED_1+lxD_2),
\end{equation}
and $y={\hat {\mathbf p}}_1 \cdot {\hat {\mathbf l}}$ and $x={\hat {\mathbf k}} \cdot {\hat {\mathbf l}}$. $N^{\prime}$ is proportional to $\omega$ so that it is infrared-convergent, and it is absorbed into Eqs.~(\ref{eq:rho1}) and (\ref{eq:rho2}). Therefore, we here only consider the contribution of $N$. With all generality, we can orient the coordinate axes so that the momentum of the charged lepton is along the $z$-axis and so that $A$ is in the first or fourth quadrant of the plane $(x,z)$. Thus, performing the sum over the photon polarization in the Coester representation and rearranging terms yields
\begin{equation}
d\Gamma_B^{\mathrm{ir}} = \frac{\alpha}{\pi} d\hat{\Omega} \frac{p_1l}{2\pi} \int_{-1}^1 dx \int_0^{2\pi} d\varphi_k \int_0^{k_m} dk \frac{k^2}{\omega}g(\theta_1) N \frac{\beta^2 (1-k^2 x^2/\omega^2)}{(\omega-\beta kx)^2}, \label{eq:b3}
\end{equation}
where $\omega^2=k^2+\lambda^2$, $k_m$ is the maximum value of the photon momentum,
\begin{equation}
k_m = \frac {{E_\nu^0}^2-(p_1-l)^2}{2\left[E_\nu^0-(p_1-l)\cos\theta_k\right]}, \label{kmax}
\end{equation}
and $g(\theta_1)$ emerges from the last $\delta$ function. It is given by
\begin{equation}
g(\theta_1) = \frac{\sin \theta_1}{a\sin \theta_1-b\cos\theta_1},
\end{equation}
$\theta_1$ is the polar angle of the decaying baryon, $a=2p_1(l+k\cos \theta_k)$, $b=2p_1k\sin \theta_k\cos \varphi_k$, $\theta_k$ and $\varphi_k$ are the polar and the azimuthal angles of the photon, respectively.

We find it convenient to consider the partition $(0,\Delta k)$, $(\Delta k,k_m)$ of the integration interval $(0,k_m)$ for $dk$, with $\Delta k$ arbitrary. Thus
\begin{eqnarray}
d\Gamma_B^{\mathrm{ir}} & = & \frac{\alpha}{\pi} d\hat{\Omega} \frac{\beta^2p_1l}{2\pi} \int_{-1}^1 dx \int_0^{2\pi} d\varphi_k \int_0^{\Delta k} dk \frac{k^2}{\omega} g(\theta_1) N \frac{1-k^2x^2/\omega^2}{(\omega-\beta kx)^2} \nonumber \\
&  & \mbox{} + \frac{\alpha}{\pi} d\hat{\Omega} \frac{\beta^2p_1l}{2\pi} \int_{-1}^1 dx \int_0^{2\pi} d\varphi_k \int_{\Delta k}^{k_m} dk \frac{k^2}{\omega} g(\theta_1) N \frac{1-k^2x^2/\omega^2}{(\omega-\beta kx)^2} \nonumber \\
& = & d\Gamma_B^{\mathrm{ir}(1)} + d\Gamma_B^{\mathrm{ir}(2)}. \label{eq:dgir}
\end{eqnarray}

We must stress that our result does not depend on $\Delta k$. The divergence is now in the first integral and because $\Delta k$ is arbitrary, we can make it slightly larger than $\lambda $, i.e., $\Delta k\gtrsim \lambda$. Then we can approximate $g(\theta_1)$ in the first integral, by allowing $k\simeq 0$. Thus,
\begin{equation}
g(\theta_1) \simeq \frac{1}{2p_1l}.
\end{equation}
Also, in $N$ we expand $y$ in powers of $k$ up to first order,
\begin{equation}
y \simeq y_{0}+f^\prime k+\mathcal{O}(k^2),
\end{equation}
where
\begin{equation}
f^\prime = \frac{1}{p_1l}\left[ E_\nu^0 + (l-p_1y_0)x-p_1(1-y_0^2)^{1/2} (1-x^2)^{1/2} \cos\theta_k\right],
\end{equation}
so that $N$ takes on the form
\begin{equation}
N \simeq \hat{A}_1^\prime+D_2p_1lf^\prime k.
\end{equation}

The first integral in Eq.~(\ref{eq:dgir}) becomes
\begin{equation}
d\Gamma_B^{\mathrm{ir}(1)} = \frac{\alpha}{\pi} d\hat{\Omega} \frac{\beta^2p_1l}{2\pi} \int_{-1}^1 dx \int_0^{2\pi} d\varphi_k \int_0^{\Delta k} dk \frac{1}{2p_1l} \frac{k^2}{\omega} (\hat{A}_1^\prime+D_2p_1lf^\prime k) \frac{1-k^2x^2/\omega^2}{(\omega-\beta kx)^2}, \label{eq:dgir1}
\end{equation}
and then the infrared divergence is finally contained in the first summand of the above equation. The second summand picks up a factor of $k$ so it is infrared-convergent and we can use $\omega=k$ in it. It will cancel away with one term of $d\Gamma_B^{\mathrm{ir}(2)}$ [see comment after Eq.~(\ref{eq:dgir2})]. A further simplification is obtained by identifying in Eq.~(\ref{eq:dgir1}) the integral of Kinoshita and Sirlin \cite{kino}
\begin{equation}
\frac{1}{2\pi} \frac{\beta^2}{2} \int_{-1}^1 dx \int_0^{2\pi} d\varphi_k \int_0^{\Delta k} dk \frac{k^2}{\omega} \frac{1-k^2x^2/\omega^2}{(\omega-\beta kx)^2} = 2\ln \left[ \frac{\Delta k}{\lambda}\right] \left( \frac{\tanh^{-1}\beta}{\beta}-1 \right) + \hat{C},
\end{equation}
where
\begin{equation}
\hat{C} = 2\ln(I_1-2) + 1 + \frac{1}{4} I_1 \left[ 2 + \ln \frac{1-\beta^2}{4} \right] + \frac{1}{\beta} L \left[\frac{2\beta}{1+\beta}\right] - \frac12 I_1 \ln \frac{1+\beta}{2}, \label{eq:C}
\end{equation}
and
\begin{equation}
I_1 = \frac{2}{\beta} \tanh^{-1}\beta.
\end{equation}

Let us now analyze the second summand in Eq.~(\ref{eq:dgir}), $d\Gamma_B^{\mathrm{ir}(2)}$. We shall change the integral over $k$ into an integral over $y$. For this purpose, we rewrite the photon momentum as
\begin{equation}
k = \frac{F}{2D},
\end{equation}
and 
\begin{equation}
N= \hat{A}_1^\prime + \frac{1}{2}D_2 F,
\end{equation}
where
\begin{eqnarray}
F = 2p_1 l(y-y_0) ,
\end{eqnarray}
and, rearranging Eq.~(\ref{eq:de}),
\begin{equation}
D = E_\nu^0+lx-p_1(xy+\sqrt{(1-y^2)(1-x^2)}\cos\varphi_k).
\end{equation}
Then we can replace
\begin{equation}
dk=\frac{dy}{2Dg(\theta_1)}.
\end{equation}
The integration limits change to $y(\Delta k)\simeq y_0+f^\prime \Delta k$ and $y(k_m)=1$. Notice that the upper limit $k_m$ of (\ref{eq:dgir}) is replaced in this variable by $y=1$.

Putting all these changes together yields
\begin{equation}
d\Gamma_B^{\mathrm{ir}(2)} = \frac{\alpha}{\pi} d\hat{\Omega} \frac{\beta^2p_1l}{2\pi} \int_{-1}^1 dx \int_0^{2\pi} d\varphi_k \int_{y_0 + f^\prime \Delta k}^1 dy\left[ \frac{\hat{A}_1^\prime}{F}+\frac{1}{2}D_2\right] \frac{1-x^2}{(1-\beta x)^2},
\end{equation}
and the integration over $y$ gives
\begin{eqnarray}
d\Gamma_B^{\mathrm{ir}(2)} & = & \frac{\alpha}{\pi} d\hat{\Omega} \frac{\beta^2}{4\pi} A_1^\prime \int_{-1}^1 dx \frac{1-x^2}{(1-\beta x)^2} \int_0^{2\pi} d\varphi_k \ln \left[ \frac{M_1(1-y_0)}{M_1f^\prime \Delta k}\right] \nonumber
\\
&  & \mbox{} + \frac{\alpha}{\pi} d\hat{\Omega} \frac{\beta^2p_1l}{2\pi} \int_{-1}^1 dx \int_0^{2\pi} d\varphi_k \frac{1}{2}D_2 \left[ (1-y_0-f^\prime \Delta k) \frac{1-x^2}{(1-\beta x)^2} \right]. \label{eq:dgir2}
\end{eqnarray}
We notice that the term proportional to $f^\prime\Delta k$ in the second summand of this latter equation cancels precisely the second summand proportional to $D_2k$ of Eq.~(\ref{eq:dgir1}) once the integration over $k$ is performed in it.

Finally, the resulting expression for $d\Gamma_B^{\mathrm{ir}}$ becomes
\begin{eqnarray}
d\Gamma_B^{\mathrm{ir}} & = & d\Gamma_B^{\mathrm{ir}(1)} + d\Gamma_B^{\mathrm{ir}(2)} \nonumber \\
& = & \frac{\alpha}{\pi} d\hat{\Omega} \hat{A}_1^\prime \left\{ 2 \left[ \frac{\tanh^{-1}\beta}{\beta}-1\right]\ln \left[ \frac{\Delta k}{\lambda}\right] + \hat{C}\right\} \nonumber \\
&  & \mbox{} + \frac{\alpha}{\pi} d\hat{\Omega} \hat{A}_1^\prime \left\{ \left[ 2\left[ \frac{\tanh^{-1}\beta }{\beta} - 1 \right] \ln \left[ \frac{M_1(1-y_0)}{\Delta k} \right] + \hat{C}_1\right] + D_2\hat{C}_2 \right\} \nonumber \\
& = & \frac{\alpha}{\pi} d\hat{\Omega} \hat{A}_1^\prime \left[ 2\left[ \frac{\tanh^{-1}\beta}{\beta} - 1 \right] \ln \left[ \frac{M_1(1-y_0)}{\lambda}\right] + \hat{C}+\hat{C}_1\right] + \frac{\alpha}{\pi} d\hat{\Omega} D_2\hat{C}_2 \nonumber \\
& = & \frac{\alpha}{\pi} d\hat{\Omega} \left[ \hat{A}_1^\prime \hat{I}_0 + D_2 \hat{C}_2\right]. \label{eq:drfinal}
\end{eqnarray}

The function $\hat{I}_0$ is defined in Eq.~(\ref{eq:drfinal}) as
\begin{equation}
\hat{I}_0 = 2\left[ \frac{\tanh^{-1}\beta}{\beta} - 1\right] \ln \left[ \frac{M_1(1-y_0)}{\lambda} \right] + \hat{C}+\hat{C}_1,
\end{equation}
where
\begin{eqnarray}
\hat{C}_1 & = & -\frac{\beta^2}{4\pi} \int_0^{2\pi} d\varphi_k \int_{-1}^1
dx\frac{1-x^2}{(1-\beta x)^2} \ln \left[ M_1 f^\prime \right] \nonumber \\
& = & -\frac12 \left\{ \ln \left| \frac{(a^+-1)(1-a^-)}{4p_1/M_1} \right|
\left[ \frac{\beta^2-1}{\beta(1-\beta x_0)} - \frac{\beta-1}{\beta} -
\frac{2}{\beta} \ln \frac{1-\beta x_0}{1+\beta} - (1+x_0) \right] \right.
\nonumber \\
&  & \mbox{} -\ln \left| \frac{(a^++1)(a^-+1)}{4p_1/M_1} \right| \left[
\frac{\beta^2-1}{\beta(1-\beta x_0)} + \frac{1+\beta}{\beta} +
\frac{2}{\beta} \ln \frac{1-\beta}{1-\beta x_0} + (1-x_0) \right] \nonumber
\\
&  & \mbox{} - 8\ln 2 + \frac{1+\beta}{\beta}\ln(1+\beta) -
\frac{1-\beta}{\beta}\ln(1-\beta) + 2 + \frac{2+\beta(1-x_0)}{1-\beta
x_0}(1-x_0)\ln(1-x_0)  \nonumber \\
&  & \mbox{} - 2 \ln(1-\beta x_0) + \frac{2-\beta(1+x_0)}{1-\beta
x_0}(1+x_0)\ln(1+x_0) + \frac{2}{\beta} \left[ L\left(
\frac{1-\beta}{1-\beta x_0} \right)
\right. \nonumber \\
&  & \mbox{}  \left. \left. - 2L\left( \frac{1-\beta}{1+\beta} \right) +
L\left( \frac{1-\beta x_0}{1+\beta} \right) + \ln \frac{\beta}{1+\beta} \ln
\frac{1-\beta}{1+\beta} - \frac12 \ln^2 \left(\frac{1-\beta x_0}{1+\beta}
\right) \right] \right\},
\end{eqnarray}
where
\begin{eqnarray}
x_0 = - \frac{l-p_1 y_0}{E_\nu^0}, \qquad \textrm{and}  \qquad a^\pm =
\frac{E_\nu^0 \mp p_1}{l},
\end{eqnarray}
and $\hat{C}$ is given in Eq.~(\ref{eq:C}). It possesses the right coefficient to exactly cancel the infrared-divergent term in its counterpart in the virtual RC, Eq.~(\ref{eq:eqocho}).

On the other hand, $\hat{C}_2$ given by 
\begin{equation}
\hat{C}_2=p_1l \frac{\beta^2}{2} (1-y_0) \int_{-1}^1 dx \frac{1-x^2}{(1-\beta x)^2},
\end{equation}
is absorbed in $\rho_2$ and $\rho_3$.

\end{document}